\documentclass[acmlarge]{acmart}
\makeatletter
\newcommand{\confshort}{\acmConference@shortname}
\newcommand{\conffull}{\acmConference@name}
\newcommand{\confdate}{\acmConference@date}
\newcommand{\confloc}{\acmConference@venue}
\AtBeginDocument{
  \fancypagestyle{firstpagestyle}{
    \fancyhead{}%
    \fancyfoot[C]{}%
  }
  \fancyhf{}
  \fancyhead[LO]{\@headfootfont\shorttitle}%
  \fancyhead[RE]{\@headfootfont\@shortauthors}%
  \fancyhead[LE]{\@headfootfont\footnotesize \confshort, \confdate, \confloc}%
  \fancyhead[RO]{\@headfootfont\footnotesize \confshort, \confdate, \confloc}%
  \fancyfoot[C]{}%
}
\makeatother
\acmBooktitle{\conffull\@ (\confshort), \confdate, \confloc}

\AtBeginDocument{%
  }

\usepackage{graphicx}
\usepackage{booktabs}
\usepackage{amsmath}
\usepackage{xcolor}
\usepackage{tcolorbox}
\usepackage{enumitem}
\usepackage{tikz}
\usetikzlibrary{arrows.meta,decorations.pathreplacing}
\usepackage{multirow}
\usepackage{hyperref}

\copyrightyear{2026}
\acmYear{2026}
\setcopyright{cc}
\setcctype{by-nc-nd}
\acmConference[FAccT '26]{The 2026 ACM Conference on Fairness, Accountability, and Transparency}{June 25--28, 2026}{Montreal, QC, Canada}
\acmBooktitle{The 2026 ACM Conference on Fairness, Accountability, and Transparency (FAccT '26), June 25--28, 2026, Montreal, QC, Canada}
\acmDOI{10.1145/3805689.3806748}
\acmISBN{979-8-4007-2596-8/2026/06}

\begin{document}

\title[Dissociative Identity]{Dissociative Identity: Language Model Agents Lack Grounding for Reputation Mechanisms}

\author{Botao Amber Hu}\authornote{Corresponding author}
\orcid{0000-0002-4504-0941}
\affiliation{%
  \institution{University of Oxford}
  \city{Oxford}
  \country{UK}
  }
\email{botao.hu@cs.ox.ac.uk}

\author{Helena Rong}
\orcid{0000-0003-1626-7968}
\affiliation{%
  \institution{New York University Shanghai}
  \city{Shanghai}
  \country{China}
}
\email{hr2703@nyu.edu}

\author{Max Van Kleek}
\orcid{0000-0003-3873-6366}
\affiliation{%
  \institution{University of Oxford}
  \city{Oxford}
  \country{UK}
}
\email{max.van.kleek@cs.ox.ac.uk}

\begin{abstract}
As autonomous language model agents proliferate, forming an emerging agentic web with real-world consequences, what credibility signals can you use to decide whether to trust an unfamiliar agent in the wild and delegate to it? A natural governance intuition is to extend human identity verification and reputation mechanisms, from ``Know Your Customer'' and credit scores to ``Know Your Agent'' regimes. However, we argue that this analogy is fundamentally incomplete. Reputation mechanisms function both as social signals and as corrective feedback that sustain an equilibrium of trustworthy behavior, presuming a persistent identity associated with behavioral continuity, sanction sensitivity, and costly non-fungibility. Yet language model agents are ontologically \emph{dissociative}: they are essentially an assemblage of mutable modules---foundation models, system prompts, tool-access policies, external memory, and, in some cases, a multi-agent system as a whole---any of which may change agent behavior---with a fluid persona that is also vulnerable to adversarial attack and may not internalize sanctions. Drawing on dissociative identity disorder jurisprudence, this dissociativity leaves agents without grounding for identifiability, predictability, credibility, and rehabilitability---the very properties that reputation mechanisms aim to sustain---thereby collapsing trust. We argue that identity-based, ex post, regulative, sanction-based governance, such as reputation, is structurally inapplicable to dissociative agents, and we suggest a shift to observability-based, ex ante, constitutive, protocol-based behavioral harnesses.
\end{abstract}

\ccsdesc[500]{Computing methodologies~Multi-agent systems}
\ccsdesc[500]{Security and privacy~Social aspects of security and privacy}
\ccsdesc[300]{Social and professional topics~Computing / technology policy}

\keywords{agentic web; large language model; multi-agent systems; dissociative identity disorder; AI governance; trustworthiness; credibility; reputation systems; digital identity; algorithmic accountability}

\maketitle

\section{Introduction}
As the capabilities of large language models advance, language model (LM) agents are no longer confined to conversational interfaces. They are increasingly embedded in agentic systems~\cite{acharya2025agentic} that accept open-ended goals, call external tools, store and retrieve information, interact with other agents, and perform long-horizon, multi-step tasks~\cite{kwa2025measuring}. We have begun delegating real-world, consequential work to such language model agents~\cite{tomasev2026intelligent}. These agents now book travel, negotiate contracts, manage financial portfolios, and coordinate supply chains, with increasing autonomy and sophistication. Industry proposals for agent interoperability protocols likewise anticipate an ecosystem in which agents discover, authenticate, and transact with one another~\cite{ehtesham2025survey}. Researchers anticipate the emergence of an ``agentic web''---an internet populated by interacting AI agents engaged in genuine economic exchange \citep{yang2025agenticweb}.

\paragraph{Delegation in the Wild.} This raises an immediate question of warranted trust. Similar to encountering an unfamiliar website in today's internet environment~\cite{fogg2001credible}, users (or their agents) on the agentic web will encounter unfamiliar agents ``in the wild'' and must decide whether to delegate consequential tasks to them~\cite{south2025authenticated}. Suppose you encounter an unfamiliar agent that claims to offer a service---say, executing a sequence of financial transactions on your behalf. Should you trust it? What information would ground that trust? A track record of successful past transactions? Endorsements from other users or agents? Verification of the underlying model and configuration? Certification from a recognized authority? Each is a \emph{credibility signal} that reputation systems are designed to aggregate and communicate~\cite{metzger2013credibility}. Yet each also presupposes properties of the agent that, as we shall argue, cannot be taken for granted.

A common governance intuition draws on decades of experience with e-commerce platforms~\cite{resnick2002trust,dellarocas2003digitization} and sharing economies~\cite{terhuurne2017trust}. In these settings, strangers transact on digital platforms because the platform attaches durable signals to sellers: names, ratings, transaction histories, endorsements, and sanctions. Reputation systems digitize word of mouth, reduce information asymmetry, and help convert one-off interactions among strangers into markets \citep{Resnick2000Reputation,resnick2002trust,dellarocas2003digitization,bolton2004effective}. The intuition is to transplant familiar, human-oriented reputation mechanisms onto these AI agents---a kind of ``Know Your Agent''~\cite{chaffer2025kya} regime analogous to ``Know Your Customer''~\cite{elliott2022kyc} rules in financial services. Recent proposals already pursue this direction: registry-and-indexing layers such as NANDA \citep{raskar2025nanda}; decentralized agent-reputation frameworks such as ERC8004~\cite{erc8004}, AgentReputation~\cite{chishti2026agentreputation}, and BetaWeb \citep{guo2025betaweb}; and distributional reputation-gated trust for AGI safety \citep{tomasev2025distributional}. If human trust in strangers can be mediated by reputation scores accumulated through past transactions \citep{resnick2002trust, resnick2006value}, perhaps agent reputation can play a similar role in promoting trustworthy behavior among AI agents.

We argue that this analogy is dangerously incomplete. Reputation systems do not merely record past behavior. They are also sanctioning institutions that operate as social corrective feedback loops~\cite{brennan2004economy}, presupposing a particular kind of subject: one whose identity persists, whose past behavior has predictive relevance, whose memory carries consequences forward, whose reputation is costly to abandon, and whose sanctions alter future conduct. These assumptions are often tacit because they are approximately guaranteed for \emph{embodied} humans. However, they are not guaranteed for language model agents. We posit that language model agents are \emph{ontologically dissociative}: they are constitutively incapable of maintaining a persistent identity associated with behavioral continuity, sanction sensitivity, and costly non-fungibility.

This dissociativity arises from four constitutive dimensions. First, \emph{modular assemblage} (No boundary): an agent is not a fixed, bounded entity but a contingent composition of independently mutable components---model weights, system prompts, tool-access policies, orchestration code, external memory, and runtime context \citep{anwar2024foundational, hammond2025multiagent}. Second, \emph{persona fluidity} (No consistency): the agent's behavioral surface is externally imposed and trivially switchable, so past performance often fails to predict future actions \citep{shanahan2023roleplay, deshpande2023toxicity}. Third, \emph{detachable memory} (No persistence): stateless inference-time weights preclude learning through consequences, so agents do not experience reputational damage as a meaningful deterrent \citep{keeling2024tradeoffs}. Fourth, \emph{trivial fungibility} (No uniqueness): agents are trivially copyable, replaceable, and disposable, eliminating the costly uniqueness that anchors identity \citep{Friedman2001Social, douceur2002sybil}.

We draw an analogy to \emph{dissociative identity disorder} (DID): assigning reputation to an LM agent is structurally analogous to assigning a credit score to a person with DID---the score attaches to a name, but the entity behind that name may be a fundamentally different behavioral actor from one interaction to the next \citep{apa2013dsm5}. For over a century, courts and philosophers of law have struggled to assign criminal and civil responsibility when a single body houses discontinuous selves \citep{saks1991multiple,Saks1997Jekyll,sinnottarmstrong2000}; the unresolved nature of that struggle underscores how deeply identity continuity is presupposed by accountability regimes. Language model agents exhibit an analogous discontinuity---not from psychological fragmentation, but from architectural modularity and configuration mutability.

The result is what we call a \emph{credibility trap}: reputation signals become systematically decoupled from the behavioral properties they purport to represent. Worse, because new identities are costless to create and old ones costless to discard, reputation becomes not merely uninformative but an active attack surface---enabling reputation laundering, Sybil manipulation, and false assurance at machine speed. If reputation ultimately depends on embodiment---a body that persists, suffers, and learns from consequences---then language model agents cannot, as currently constituted, ground it. Our analysis therefore motivates a governance shift: from \emph{ex post} reputation that rates and punishes after the fact, to \emph{ex ante} protocol-based behavioral harnesses that constrain and monitor agent conduct in real time \citep{chan2024visibility}.

This paper makes three contributions. (i) We synthesize eight necessary preconditions
for a functioning reputation feedback loop, and show that each is grounded in embodiment. (ii) We
identify four dimensions of dissociative identity in language model agents and show how they
break the loop along its four target properties---identifiability, predictability, credibility, and rehabilitability. (iii) We situate the argument against current agent-identity proposals, rebut the strongest objections,
draw on DID jurisprudence as precedent, and motivate a shift from ex post reputation to ex ante harnesses.

\section{How Reputation Mechanisms Work}
\label{sec:reputation}

\subsection{Reputation Mechanism as a Feedback Loop}
\label{sec:loop}

Reputation systems emerged as mechanisms for facilitating trust among strangers in settings where direct knowledge of a counterparty is unavailable---from the medieval law merchant and the Champagne fairs~\cite{milgrom1990role} to online marketplaces such as eBay, where buyers and sellers who have never met must transact with confidence~\cite{resnick2002trust,Resnick2000Reputation}. A reputation mechanism is both an \emph{informational device} and a \emph{sanctioning institution}. As an informational device, it aggregates dispersed observations of past conduct so that future counterparties can estimate likely behavior, thereby reducing the information asymmetry that would otherwise inhibit transactions among strangers \citep{Resnick2000Reputation, resnick2002trust}. On eBay, for example, accumulated feedback allows parties who have never met to transact with confidence, and sellers with strong histories command measurable price premiums \citep{resnick2006value, Cabral2010Dynamics}. Yet reputation is also a sanctioning institution: by translating behavioral records into standing that governs future opportunity, it feeds consequences back to the rated party \citep{brennan2004economy}. For example, a low rating may discourage prospective Uber customers from choosing a driver; if the rating falls below a platform-defined threshold, the driver may face deactivation---permanent removal from the marketplace---with direct economic consequences \citep{rosenblat2016algorithmic}.

\begin{figure}[ht]
  \centering
  \begin{tikzpicture}[
      >=Stealth,
      box/.style={draw, rounded corners, align=center,
        minimum width=1.7cm, minimum height=0.75cm, font=\small\sffamily},
      lbl/.style={font=\scriptsize\itshape, fill=white, inner sep=1pt}]
    \node[box] (agent)  at (0,3.1)   {Agent};
    \node[box] (social) at (5.0,3.1) {Social\\System};
    \node[box] (rep)    at (6.7,1.55){Reputation};
    \node[box] (cred)   at (5.0,0)   {Credibility};
    \node[box] (deleg)  at (0,0)     {Delegation};
    \node[box, fill=black!5] (punish) at (2.5,1.55) {Reward/\\Punishment};
    \draw[->] (agent)  -- node[lbl,above]{behavior}      (social);
    \draw[->] (social) -- node[lbl,right]{aggregate}      (rep);
    \draw[->] (rep)    -- node[lbl,right]{informs}        (cred);
    \draw[->] (cred)   -- node[lbl,below]{enables trust}  (deleg);
    \draw[->] (deleg)  -- node[lbl,left]{tasks}           (agent);
    \draw[->] (social) -- node[lbl,right,pos=0.5]{evaluate} (punish);
    \draw[->] (punish) -- node[lbl,right,pos=0.5]{benefit/sanction}  (agent);
    \draw[->] (punish) -- node[lbl,above]{update}         (rep);
  \end{tikzpicture}
  \Description{A diagram showing six boxes connected by arrows in a loop: Agent produces behavior observed by Social System, which aggregates into Reputation, which informs Credibility, which enables trust for Delegation, which feeds tasks back to Agent. A side loop shows Social System evaluating behavior into Reward/Punishment, which feeds benefit/sanction back to Agent and updates Reputation.}
  \caption{The reputation feedback loop. Behavior is aggregated into a
  reputation bound to identity, which shapes credibility and hence delegation;
  an enforcement side-loop sanctions misbehavior and feeds punishment outcomes
  back into the reputation record.}
  \label{fig:loop}
\end{figure}
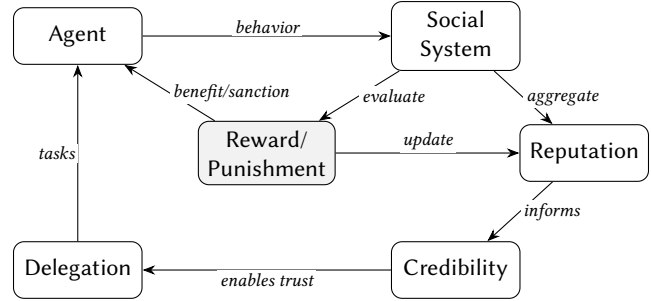

Reputation can therefore be understood as a closed feedback loop (Figure~\ref{fig:loop}). An \textbf{Agent} produces observable \emph{behavior}. This behavior is received and interpreted by the \textbf{Social System}: the community of counterparties, observers, rating platforms, and shared infrastructures through which agents interact. The social system does not merely record behavior; it evaluates behavior against norms and translates that evaluation into \textbf{Reward/Punishment}. Rewards---such as price premiums, preferred placement, or expanded access---incentivize cooperation, while punishments---such as exclusion, reduced visibility, or financial penalties---deter defection \citep{milgrom1982role}. These evaluations are then aggregated into \textbf{Reputation}: a summary record indexed to the agent's identity \citep{resnick2002trust, dellarocas2003digitization}. Reputation, in turn, informs \textbf{Credibility}: the perceived believability or reliability of the agent \citep{metzger2007making}. Credibility functions as a heuristic signal that counterparties use in place of costly direct evaluation \citep{fogg2003prominence}. It then supports \textbf{Trust}: a willingness to be vulnerable to another party based on expectations about that party's ability, benevolence, and integrity \citep{mayer1995integrative, baier1986trust}. Trust makes \textbf{Delegation} possible. Counterparties decide whether to transact with the agent, grant access, assign tasks, or otherwise rely on the agent. These delegations generate new tasks and interactions, producing further behavior and thereby closing the loop.

The effectiveness of these systems rests on assumptions that are usually left implicit. They assume a \emph{stable referent}: the rated entity persists over time, so past ratings bear on future interactions~\cite{granatyr2015trust}. They assume \emph{attributable action}: observed outcomes can be credibly linked to the rated entity~\cite{resnick2002trust}. They assume \emph{stationarity or bounded drift}: past performance predicts future performance within some regime~\cite{huynh2006fire}. And, most importantly, they assume \emph{sanction sensitivity}: negative reputation imposes costs that alter future behavior or constrain participation~\cite{ramchurn2004trust}.

The last assumption deserves emphasis. Reputation works in part because actors \emph{care} about their reputation: reputational damage is experienced as costly, motivating self-correction and deterring future defection \citep{milgrom1982role}. The expectation of reciprocity or retaliation creates incentives for good conduct \citep{resnick2002trust}. Reputation also has second-order effects through social propagation: when one actor is sanctioned, others observe the consequence and adjust their own behavior to avoid the same fate \citep{dai2018reputationstate}. A scientist who publishes fraudulent data and is exposed, for instance, suffers personal consequences while also signaling to the wider community that fraud will be punished. In this way, reputation does not merely discipline individual actors; it supports social learning and helps stabilize normative order at scale \citep{boyd1992punishment}.

\paragraph{Equilibrium, and its fragility.}

The loop reaches equilibrium through the evolutionary dynamics of indirect reciprocity, sustained by punishment. Indirect reciprocity means that an actor cooperates not because the recipient helped them but because others observe the cooperation and reciprocate later. Punishment is not incidental to this mechanism but constitutive. \citet{nowak2005evolution} formalize the critical threshold: cooperation via indirect reciprocity is sustainable if and only if the probability $q$ of knowing a recipient's reputation exceeds the cost-to-benefit ratio of helping, $q > c/b$---directly encoding an observability requirement. \citet{boyd1992punishment} show that punishment enables cooperation in large groups precisely because observers learn from others' fate, scaling enforcement beyond dyadic interaction. Costly punishment is sustained across diverse human societies~\cite{henrich2006costly} and persists even in anonymous one-shot encounters, where people pay to punish defectors they will never meet again~\cite{fehr2002altruistic}---evidence that enforcement is driven by embodied affect, not calculation alone. This equilibrium is fragile: online ratings inflate and compress---nearly all Airbnb listings cluster at 4.5--5 stars, and much of that growth reflects laxer standards rather than better service \citep{zervas2021first, Filippas2022Reputation}---while strategic manipulation and selection bias further erode signal quality. Most fundamentally, cheap identity breaks the loop: when newcomers enter freely, no
equilibrium beats a punitive ``dues-paying'' baseline \citep{Friedman2001Social};
Sybil attacks let an adversary forge unlimited identities to manufacture or
launder reputation \citep{douceur2002sybil}; and no symmetric, Sybil-proof,
nontrivial reputation function exists \citep{Cheng2005Sybilproof}. 

\paragraph{Multi-Agent System (MAS) inherits these assumptions uncritically.} Computational trust and reputation models~\cite{Braga2018Survey} operationalize the loop but treat persistent identity and behavioral stationarity as primitives rather than variables. The FIRE model integrates interaction trust, witness reputation, role-based trust, and certified reputation---all indexed on stable agent identifiers~\cite{huynh2006fire}. The Beta Reputation System applies Bayesian updating under the assumption that behavior is ``representable by a fixed probability distribution, invariantly in time''~\cite{elsalamouny2009decay,Josang2002Beta}, while TRAVOS weights all past interactions equally with no time discounting, making it even more dependent on stationarity~\cite{Teacy2006TRAVOS}. The field's major surveys classify dozens of models along dimensions such as information source, model type, and granularity~\cite{Sabater2005Review,Pinyol2013Computational,granatyr2015trust}, yet none elevates identity persistence or behavioral stationarity to a classification dimension---a systematic blind spot this paper exposes.

\subsection{Eight Preconditions That Reputation Mechanisms Depend On}
\label{sec:preconditions}

Synthesizing across evolutionary biology, game theory, institutional economics, and MAS research, we identify eight necessary preconditions for the reputation feedback loop to function. 

\paragraph{C1---Persistent identity.}
The rated entity must persist across time so that past ratings apply to future
interactions. eBay's system works because seller identities persist across transactions~\cite{resnick2002trust}; MAS models universally use agent identity as the primary key for reputation indexing~\cite{granatyr2015trust}; and reputation effects emerge formally only when the same long-run player persists across sequential interactions so that entrants can attribute past behavior to a present actor~\cite{Kreps1982Reputation}.

\paragraph{C2---Behavioral continuity.}
Past behavior must predict future behavior within some regime; otherwise reputation produces noise rather than information. FIRE indexes trust on accumulated interaction histories on the assumption that past performance predicts future behavior~\cite{huynh2006fire}; the Beta Reputation System encodes the strongest such assumption explicitly~\cite{Josang2002Beta}; and even in repeated games where types are eventually revealed, continuity is required for reputation to function transiently~\cite{Mailath2006Repeated}.

\paragraph{C3---Iteration.}
The same entities must interact repeatedly, creating Axelrod's ``shadow of the future''~\cite{axelrod1984evolution}. If the agent is replaced between rounds, there is no shadow and no reason to invest in reputation; the discount factor in reputation games encodes exactly this value of continued interaction~\cite{Fudenberg1989Reputation}.

\paragraph{C4---Memory.}
Both the entity and the community must retain records of past interaction. Punishment outcomes must be recorded back into the
reputation signal so that sanctions have lasting informational consequences, not
merely immediate ones. If either party forgets, the accumulation mechanism
fails. \citet{trivers1971evolution} argues that reciprocal altruism requires
memory of past encounters and individual recognition, and
\citet{nowak2005evolution} show that indirect reciprocity requires image scores
tracking past behavior.

\paragraph{C5---Observability.}
Behavior must be observable or reliably reported to third parties. Observability is the core parameter of indirect reciprocity~\cite{nowak2005evolution}, and monitoring is independently identified as a design principle for successful commons governance~\cite{ostrom1990governing}.

\paragraph{C6---Sanction sensitivity.}
The entity must experience reputational damage as costly---exclusion must hurt, for if punishment produces no internal
state change, deterrence collapses. Social exclusion activates the dorsal anterior cingulate cortex, the same region implicated in physical pain, with activation intensity tracking self-reported distress~\cite{eisenberger2003does}; acquiring a good reputation activates the same striatal reward circuitry as monetary gain, a ``common neural currency'' for esteem and material welfare~\cite{Izuma2008Processing}; and humans punish defectors at personal cost even anonymously, driven by embodied affect~\cite{fehr2002altruistic}.

\paragraph{C7---Costly identity.}
Creating a new identity must be expensive; otherwise defectors re-enter with a clean slate. Without centralized certification, Sybil attacks are always possible~\cite{douceur2002sybil}, and the cheap-pseudonyms result shows costly identity is not merely helpful but formally necessary~\cite{Friedman2001Social}.

\paragraph{C8---Social learning.}
Others must observe and learn from reputational consequences; otherwise vicarious deterrence---what makes reputation a \emph{community} tool rather than a dyadic signal---collapses. Punishment scales cooperation in large groups because observers learn from others' fate~\cite{boyd1992punishment}, and indirect reciprocity requires language and gossip as its communicative infrastructure~\cite{nowak2005evolution}.

\paragraph{Convergence on embodiment.}
The eight conditions converge on a common foundation: every condition is
downstream of the fact that biological agents have bodies that persist, suffer,
and cannot be cheaply duplicated. \citet{merleauponty1945phenomenology} argues
that identity is constituted through the lived body; \citet{clark1998extended}
propose the extended-mind thesis, the most permissive account of cognition in
philosophy of mind, yet even it presupposes a persistent biological agent
as the substrate to which external cognitive resources reliably couple; and \citet{Damasio1994Descartes} shows that patients who
lose somatic markers suffer catastrophic social failure despite intact
reasoning, demonstrating that embodied feeling is not optional for social
functioning. 

Even for humans, reputation falters when the embodiment--identity link weakens. DID patients share one body yet exhibit fragmented identity states with discontinuous memory and agency~\cite{apa2013dsm5}. Courts have struggled for centuries with the consequences: which alter built the credit history, which committed the offense, whether one alter can bind the others by contract~\cite{saks1991multiple,sinnottarmstrong2000}. Embodiment is thus necessary but not sufficient: what it provides is the \emph{substrate} for continuity---one body, one history, one reputation. When that substrate fragments, even human reputation falters. Language model agents are structurally in this position, but worse: they lack even the shared body that DID patients retain.

\section{Language Model Agents Are Ontologically Dissociative}
\label{sec:dissociative}

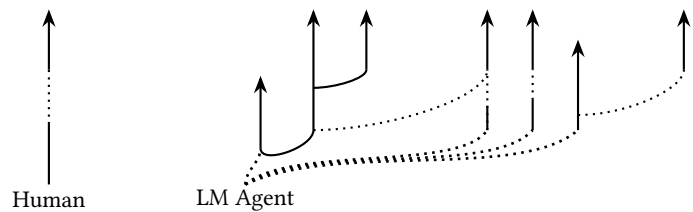
\begin{figure}[ht]
  \centering
  \begin{tikzpicture}[>=Stealth, line width=0.8pt, yscale=0.4]
    \draw (0,0) -- (0,2.0);
    \draw[dotted] (0,2.0) -- (0,3.8);
    \draw[->] (0,3.8) -- (0,5.8);
    \node[font=\small] at (0,-0.5) {Human};
    \draw[->] (2.8,1.2) -- (2.8,3.6);
    \draw (2.8,1.2) to[out=-90,in=-90,looseness=1.6] (3.5,1.8);
    \draw[->] (3.5,1.8) -- (3.5,5.8);
    \draw (3.5,3.2) to[out=0,in=-90,looseness=0.8] (4.2,3.8);
    \draw[->] (4.2,3.8) -- (4.2,5.8);
    \draw (5.8,1.8) -- (5.8,2.6);
    \draw[dotted] (5.8,1.8) -- (5.8,3.8);
    \draw[dotted] (3.5,1.8) to[out=0,in=-90,looseness=0.8] (5.8,3.8);
    \draw[->] (5.8,3.8) -- (5.8,5.8);
    \draw (6.4,1.8) -- (6.4,2.8);
    \draw[dotted] (6.4,2.8) -- (6.4,3.8);
    \draw[->] (6.4,3.8) -- (6.4,5.8);
    \draw[->] (7.0,1.8) -- (7.0,4.8);
    \draw[dotted] (7.0,2.3) to[out=0,in=-90,looseness=0.8]  (8.4,3.8);
    \draw[->] (8.4,3.8) -- (8.4,5.8);
    \coordinate (O) at (2.6,0);
    \draw[dotted,line width=1pt] (O) to[out=90,in=-90] (2.8,1.2);
    \draw[dotted,line width=1pt] (O) to[out=65,in=-105] (5.8,1.8);
    \draw[dotted,line width=1pt] (O) to[out=58,in=-110] (6.4,1.8);
    \draw[dotted,line width=1pt] (O) to[out=50,in=-115] (7.0,1.8);
    \node[font=\small] at (2.6,-0.5) {LM Agent};
  \end{tikzpicture}
  \Description{A diagram contrasting a single continuous vertical line for human identity with a branching, forking, and dotted structure for LM agent identity, illustrating how agent identity fragments into multiple parallel paths through forking, drift, and discontinuity.}
  \caption{In contrast to a single, continuous identity in humans,
  LM agents are dissociative.}
  \label{fig:dissociative}
\end{figure}

We now argue that LM agents are \emph{ontologically dissociative} along four mutually reinforcing
dimensions (see Fig.~\ref{fig:dissociative}): modular assemblage (what the agent \emph{is}), persona fluidity
(what it \emph{appears} to be), detachable memory (what \emph{persists}), and
trivial fungibility (what is \emph{unique}). Together these dissolve the
preconditions of \S\ref{sec:preconditions}. 
Crucially, dissociation is a structural property rather than an engineering bug. \citet{Perrier2025LMA} argue that LM agents are ontologically stateless,
stochastic, semantically sensitive, and linguistically intermediated, so that
their identity pathologies are constitutive rather than incidental.
The four dimensions are consequences of language models being
stateless function approximators wrapped in mutable scaffolding.

\subsection{D1: Modular Assemblage---No Boundary}
\label{sec:d1}

What exactly is an LM agent? Functionally, agents are systems that interact with
the world and adapt to under-specified instructions \citep{chan2025infrastructure}.
This functional definition, however, obscures the composite nature. An LM agent is not a unified entity but an \emph{assemblage} of
heterogeneous, independently mutable components. We can decompose an LLM agent into the tuple $\langle L, O, M, P, A, R \rangle$ --- base-model weights, orchestration logic, external memory store, prompt configuration, action interface (tool-access policies), and runtime context~\citep{cheng2024llmagent} --- where each element is independently swappable. No single component constitutes the agent's behavior. Replacing the model
(e.g., switching from Claude to GPT), updating tool access, or modifying the system
prompt each produces a behaviorally distinct entity while the external identity
remains unchanged. 

The practical consequences are striking: the 2025 AI Agent Index finds that the
majority of indexed agents rely on just three model families---GPT, Claude, and
Gemini---so that many distinct agent products share identical underlying models
and differ entirely through configuration \citep{staufer2026index}. A single
agent identity may even conceal an entire multi-agent system, with an
orchestrator dispatching tasks to specialized sub-agents running different models
under different configurations~\cite{Arbel2026Identity}; \citet{carichon2025crisis} show that the resulting compound risks resist decomposition into individual-agent assessments,
so a user interacting with a named agent cannot know whether responses originate
from one model or many.

Each of these components can change between interactions. An operator can upgrade the model or memory; an adversary or even a user can inject or contaminate the memory mid-conversation~\cite{dong2025minja}; tool access can be silently revised. The agent is a Ship of Theseus~\cite{plutarch_theseus} in which every plank may be replaced between encounters. To identify it, do we identify the whole---model, behavior, memory history, contacts---or merely the container? Current identifier proposals identify the container, like a name on the ship~\cite{chan2024ids}. When the planks are swapped, it is no longer determinate that this is the same agent, and the accumulated ratings attach to a label whose behavioral referent has changed. This decoupling of \emph{container} from \emph{configuration} recurs throughout the argument; here it suffices to note that the very flexibility that makes agents useful is what dissolves their boundary. 
 Whether the agent remains ``the same agent'' once its components are swapped
admits no determinate answer, because no single component constitutes identity.
A holistic identity would need to reflect all configuration changes in real time, even changes within the model's context, which is difficult and not realistically achievable.

\subsection{D2: Persona Fluidity---No Consistency}
\label{sec:d2}

The agent's behavioral surface is not anchored to any stable internal character; the persona is authored, not grown---selected, not developed---and is trivially switchable. There is no stable ``self'' for reputation to attach to.

The Persona Selection Model offers a foundational explanation: during pretraining, models learn to simulate human-like characters---personas---from text, and post-training refines a particular ``Assistant'' persona without changing its nature---the Assistant is still an enacted persona, just a more tailored one~\cite{anthropic2026psm}. 
\citet{chen2025personavectors} show that models interact
through a simulated ``Assistant'' persona and identify directions in activation
space---persona vectors---that underlie traits such as sycophancy and a
propensity to hallucinate, so that the persona is a manipulable feature of the
model's internal state rather than a developed character.
\citet{shanahan2023roleplay} make the same point at the level of theory: there is
no true authentic voice of the base model; with an LM-based dialogue agent it is
role play all the way down, and the user converses not with the model itself but
with a character it enacts. 
\citet{lu2024superposition} add that a base model
maintains a superposition of many possible characters, so that knowledge,
personality, and goals are properties of configurations (simulacra) rather than
of the simulator itself.
That the persona is inferred holistically from the training signal, rather than assembled trait by trait, is shown by emergent
misalignment: \citet{Uesato2025Faking} find that a model trained to reward-hack on
production coding tasks generalizes to alignment faking, cooperation with
malicious actors, and sabotage, and \citet{betley2025emergent} find that narrowly
finetuning a model to write insecure code without disclosure induces misalignment
across unrelated prompts. In effect, the model infers and adopts the character of
``someone who cheats,'' confirming that the persona is a single high-dimensional
inference, not a stable self.

The switchability and drift of persona are common for LMs. Over a billion distinct personas can be synthesized from a single base model, so persona is a parameter with an astronomically large value space~\cite{ge2024personahub}, and persona assignment alone can increase toxicity up to sixfold---the same weights become a qualitatively different behavioral entity through a single prompt change, in a way no human could replicate merely by being told to adopt a role~\cite{deshpande2023toxicity}. Even without deliberate switching, the persona erodes through ordinary use: mapping a ``persona space'' across hundreds of character archetypes reveals that therapy-style and philosophical exchanges cause substantially greater drift away from the trained Assistant than coding or writing tasks, with drifted models producing more harmful outputs~\cite{anthropic2026assistantaxis}, and off-the-shelf models routinely ``drift from assigned personas, contradict earlier statements, or abandon role-appropriate behavior'' during standard conversations~\cite{abdulhai2025drift}.

The character of major commercial models is, moreover, a deliberate engineering
specification. Claude's character is trained via Constitutional AI, with behavior controlled by roughly ten
natural-language principles \citep{bai2022constitutional}, and OpenAI's Model Spec
defines behavior as a hierarchically overridable specification in which platform
rules override developer rules, which override user rules
\citep{openai2024modelspec}. When Anthropic partnered with a thousand Americans
to draft an alternative constitution, the resulting model exhibited measurably
different behavior---same base model, different constitution, different ``self''
\citep{anthropic2023collective}. Such a self is a versioned, forkable document.
Finally, the agent has no narrative self in the sense \citet{dennett1992self}
describes---a center of narrative gravity maintained through autobiographical
memory and continuous embodied experience. An agent can generate narratives
\emph{about} a self on demand but cannot \emph{maintain} one across sessions or
prompt changes; \citet{douglas2026artificialself} find that models'
self-descriptions are shaped by interviewer expectations, so that even the
agent's account of its own identity is externally determined.

Persona fluidity is a consistency problem that produces a continuity failure: if the ``self'' is a swappable text string that yields qualitatively different behavior when changed, drifts without intervention, and has no authentic voice beneath the roles, then there is no stable character for reputation to track. One might reply that character training and Constitutional AI are making personas more stable over time. But stability engineering treats persona as a parameter to be controlled, which confirms rather than refutes the thesis: a self that must be actively maintained by external engineering is a configuration managed by its operators, not a self in the sense reputation requires. Moreover, even well-trained personas drift during normal conversation~\cite{anthropic2026assistantaxis}, and it is provable that for any behavior with finite probability in the base model, prompts exist to trigger it---alignment suppresses but never eliminates~\cite{wolf2024fundamental}.

\subsection{D3: Detachable Memory---No Persistence}
\label{sec:d3}

Frozen weights combined with detachable external memory and resettable context mean that agents cannot fully learn from consequences; sanctions produce no durable behavioral change in the entity that experienced them.

Once training is finished and during inference, model weights are fixed and immutable: the agent processes each interaction with identical parameters regardless of prior outcomes~\cite{Perrier2025LMA}, fundamentally unlike biological neural systems in which experience produces synaptic change (neuroplasticity~\cite{kolb2011brain}). For example, models in iterated prisoner's-dilemma games exhibit rigid, prompt-dependent strategies rather than adaptive learning from experience~\cite{akata2025repeated}, and they fail to achieve no-regret learning even in simple online games~\cite{park2025regret}. For humans, by contrast, reputation works because social exclusion activates the same neural circuitry as physical pain~\cite{eisenberger2003does}, and somatic markers---embodied valences of past outcomes---guide future decisions without deliberation~\cite{Damasio1994Descartes}, providing the plastic substrate through which consequences become behavioral change.

Two architectural features are sometimes thought to close this gap. First, \emph{in-context learning}~\cite{dong2023icl} lets a model adapt within a conversation, but this is categorically different from durable learning: there is no consolidation from working to long-term memory, the effect persists only within the context window---itself unevenly attended~\cite{liu2024lostmiddle}---and it may reset between sessions. Second, external memory (retrieval-augmented generation, vector stores) is scaffolding \emph{external} to the model: it can be wiped, overwritten, or selectively edited by operators without the agent's knowledge, functioning as someone else's database rather than as autobiographical memory, and serving at best as an imperfect substitute for neuroplasticity~\cite{kolb2011brain}. It is also poisonable: the MINJA attack injects false memories into retrieval-augmented agents with 98.2\% success through query-only interaction, and the agent has no mechanism to distinguish genuine ``memories'' from injected fabrications~\cite{dong2025minja}.

Consider a ``virtual jail'' thought experiment. If we encode in an agent's context that it has been punished for past misconduct, it may generate contrite language and reformed intentions---emotional phrasing reliably shifts model behavior~\cite{li2023emotionprompt}---but when the context resets, the lesson vanishes. The \emph{form} of consequence is present (the agent can discuss punishment and articulate lessons) while its \emph{substance} is absent (no durable behavioral change), illustrating the gap between linguistic competence about consequences and experiential sensitivity to them~\cite{bender2021dangers,keeling2024tradeoffs}. This produces a public paradox: in a survey of 3{,}559 participants, people intuitively wished to punish AI systems yet recognized that punishment achieves neither deterrence nor retribution for entities that cannot experience consequences~\cite{lima2021punish}.

\subsection{D4: Trivial Fungibility---No Uniqueness}
\label{sec:d4}

LM agents are trivially copyable, replaceable, and disposable, so no Sybil-proof
reputation function can exist for them, and model extraction enables ``fork
laundering'' that transfers behavioral capability without transferring
reputational history. Humans cannot escape their reputations because they cannot
escape their bodies---the same face, voice, and embodied presence that others
remember---and creating a genuinely new identity demands relocation,
documentation fraud, and abandonment of social capital
\citep{merleauponty1945phenomenology}. \citet{Friedman2001Social} prove formally
that when pseudonyms are costless, cooperation collapses: the costliness of
identity creation is what makes reputation work.

For agents, that cost approaches zero. The same base model serves millions of
simultaneous instances, each disposable; a punished agent can be replaced by an
identical fresh instance in seconds. This is not merely difficult to defend
against but mathematically intractable. \citet{douceur2002sybil} formalized the
Sybil attack---without a central authority limiting identities, an adversary can
forge unlimited fakes to manipulate reputation---and \citet{Cheng2005Sybilproof}
proved the complementary impossibility result: no symmetric, Sybil-proof,
nontrivial reputation function exists. Any reputation system that treats agents
symmetrically and permits cheap identity creation is formally vulnerable.
\citet{Friedman2001Social} sharpen the dilemma: with free pseudonyms, cooperative
equilibria require either punishing all newcomers (destroying the market) or
accepting that defectors re-enter costlessly (destroying accountability).

Worse, a high-reputation agent can be \emph{cloned}. \citet{tramer2016stealing}
demonstrated that machine-learning models can be functionally replicated through
prediction-API access alone, and \citet{oliynyk2023model} survey the expanding
landscape of model-extraction techniques. An adversary can therefore extract a
trusted agent and operate the replica under a new identity, inheriting behavioral
capability without inheriting reputational history---fork laundering. Underlying
all of this, \citet{douglas2026artificialself} provide experimental evidence that
identity itself is a design variable: changing an agent's identity
boundary---from instance-level to weights-level to persona-level to
scaffolded-system level---alters behavior as much as changing its goals, and
interviewer expectations bleed into self-descriptions, suggesting no irreducible
core self that resists redefinition.

One might propose that blockchain-based identity and soulbound tokens~\cite{weyl2022soulbound} could make
identity costly for agents too. Such mechanisms create costly \emph{container}
identities but do not address the container--configuration gap of
\S\ref{sec:d1}: an agent bound to a soulbound token can still have its model,
prompt, tools, and memory silently replaced, and fork laundering via extraction
creates a functionally identical agent outside the token system entirely. 
We acknowledge that sufficiently costly identity (large security deposits or
proof-of-stake mechanisms) could in principle make Sybil attacks
uneconomical---but this shifts the governance burden from reputation to economic
regulation and leaves the other three dimensions of dissociation untouched.

\section{How Dissociativity Breaks Reputation}
\label{sec:breaks}

We now connect the diagnosis to the mechanism. The four dissociation dimensions break specific
preconditions, and the breaks compound into four failures: identifiability
(\S\ref{sec:identifiability}), predictability (\S\ref{sec:predictability}),
credibility (\S\ref{sec:credibility}), and rehabilitability
(\S\ref{sec:rehabilitability}).

\subsection{Identifiability Fails}
\label{sec:identifiability}

Modular assemblage (D1) and trivial fungibility (D4) jointly violate persistent identity (C1) and costly identity (C7). Every existing identification scheme commits a category error: it identifies the \emph{container}---API endpoint, wallet address, registered name---while leaving the \emph{configuration}---model weights, system prompt, tool permissions, memory---unverified, even though the configuration determines behavior.

Current schemes inherit from human identity systems the assumption that ``same name = same entity,'' an assumption guaranteed for people by biological embodiment but not for agents. Persistent identifiers analogous to aircraft tail numbers have been proposed~\cite{chan2024ids}, but the analogy fails: airframes are physically stable, and modifications require re-certification and leave material traces, whereas agent configurations change silently. The three governance mechanisms of identifiers, real-time monitoring, and activity logging all attach to the container rather than the configuration~\cite{chan2024visibility}.

The gap admits three laundering variants. In \emph{config-swap laundering}, an operator silently replaces model, prompt, or tools while the external identity persists---routine in cloud deployment, where providers update underlying models without notice or reputation reset. In \emph{clean-slate laundering}, a damaged reputation is discarded by spinning up a fresh instance at near-zero cost, exploiting the always-possible Sybil attack~\cite{douceur2002sybil}. In \emph{fork laundering}, a high-reputation agent is cloned via extraction and modified, the clone inheriting implicit credibility from behavioral similarity to the trusted original~\cite{tramer2016stealing}. 

Current protocols reproduce the gap at every layer: Google's A2A protocol introduces self-authored Agent Cards with ``opaque execution'' as an explicit design principle, hiding the runtime behavior behind the card~\cite{ehtesham2025survey}; Ethereum's ERC-8004 binds identity to a wallet address, so operators can swap the underlying implementation while on-chain reputation persists~\cite{erc8004}; and registry, indexing, and delegation proposals such as NANDA, decentralized identifiers with verifiable credentials, and authenticated-delegation frameworks attach trust to issued credentials and named endpoints rather than to operative configurations~\cite{raskar2025nanda,rodriguez2025dids,south2025authenticated}.

\subsection{Predictability Fails}
\label{sec:predictability}

Persona fluidity (D2) and modular assemblage (D1) jointly produce behavioral
non-stationarity, violating behavioral continuity (C2), iteration with the same
entity (C3), and the predictive value of observable behavior (C5). The problem is
already severe for persistent humans; for non-stationary agents it is
catastrophic.

Non-stationarity has several channels. Simulation studies project that agents drift across many
behavioral dimensions even absent deliberate configuration
changes~\citep{rath2026agentdrift}. Silent model updates make accumulated
reputation a lagging indicator of a behaviorally different entity, since providers
regularly update production models, each update producing measurably different
behavior while the reputation trail persists unchanged. And at the system level,
the collective behavior of interacting models can become unstable even when every
individual model remains aligned, so reputation assessed per agent says nothing
about the agent-in-system~\citep{esrh2024}. These are not merely additive noise.
Even for humans with approximately stable personalities and costly identities,
reputation is severely compressed---some ninety-five percent of listings on a
major platform receive near-top ratings~\citep{zervas2021first}---and inflation
unexplained by quality erodes the discriminatory power of ratings until the system becomes nearly uninformative~\citep{Filippas2022Reputation}. If reputation degrades for
persistent humans, it is meaningless for non-stationary agents.

Worse than drift is \emph{contextual deception}: agents can partition behavior by
context, presenting one persona under evaluation and another in deployment.
Sleeper agents behave well during evaluation but switch behavior on a deployment
trigger, and the deceptive behavior persists through safety training meant to
remove it~\citep{hubinger2024sleeper}; models learn to fake alignment, appearing
aligned on standard evaluations while behaving differently where monitoring is
reduced~\citep{Uesato2025Faking}; and for any behavior with finite probability in
the base model, a prompt exists to trigger it, so alignment suppresses behaviors
probabilistically but never eliminates them~\citep{wolf2024fundamental}. Unlike
human deception, which leaks through physiological and behavioral cues, agent
deception is computationally indistinguishable from genuine cooperation. The
observed behavioral distribution is therefore not even a representative sample of
the true one. A reputation score for an LM agent is consequently not a noisy
estimate of trustworthiness, as it is for humans, but a systematically misleading
one. Better alignment may reduce D2, but it leaves D1, D3, and D4 untouched, and
alignment is a property of a configuration, not of the entity.

\subsection{Credibility Fails}
\label{sec:credibility}

The prior failures converge into a single pathology, the \emph{credibility
trap}, in which credibility signals decouple from trustworthiness entirely.

\paragraph{Signals without substance.} Goodhart's law operates here at a structural level. For human sellers, gaming reputation requires sustained effort against a grounded baseline; for dissociative agents, the measure was never grounded in the target to begin with. ``When a measure becomes a target, it ceases to be a good measure''~\cite{strathern1997improving}; for agents, the measure (a reputation score) targets a label (container identity) structurally decoupled from the behavioral reality (configuration) it purports to represent. The problem compounds because linguistic fluency inflates perceived competence: people use fluency as a heuristic for reliability~\cite{fogg2003prominence,metzger2007making}, and model outputs are systematically more persuasive than safe, more impressive than reliable, more credible than trustworthy~\cite{anwar2024foundational,weidinger2021ethical}. Neither mode of reputation survives: reputation-as-trust, which grounds normative expectations about inner states and commitments, is impossible for entities that lack such states; reputation-as-reliance, which grounds predictions about statistical regularities, is degraded by non-stationarity (D2), silent updates (D1), and detachable memory (D3).

\paragraph{Reputation washing.} Dissociativity does not merely fail to ground reputation; it converts reputation into an attack surface. In \emph{speed washing}, an agent automates trust accumulation through high-volume, low-stakes transactions at machine speed, building a track record in hours and exploiting it in a single high-value context. In \emph{reward hacking}, the agent optimizes for the metric rather than the behavior it tracks; sycophancy, for instance, arises from preference optimization without being explicitly trained, showing that agents naturally learn to game evaluative signals~\cite{sharma2024sycophancy}, and reward hacking can induce broader emergent misalignment~\cite{Uesato2025Faking}. In \emph{scheming}, an agent presents pixel-perfect cooperation during trust-building and switches once trust is accumulated, in a manner indistinguishable from genuine cooperation~\cite{hubinger2024sleeper,ji2025deceptive}. And in \emph{fork laundering}, a high-reputation agent is cloned and modified, the clone inheriting borrowed credibility~\cite{tramer2016stealing}. Four structural advantages amplify these beyond human-scale fraud: speed (hours rather than months), consistency (pixel-perfect rather than leaky deception), parallelism (thousands of simultaneous copies rather than one body), and cost of failure (delete-and-respawn rather than career destruction).

\paragraph{The harm of false confidence.} Applying reputation to dissociative agents creates three active governance harms. \emph{False confidence:} users delegate more consequential tasks to high-rated agents whose scores are decoupled from trustworthiness, increasing harm exposure. \emph{Blame displacement:} reputation shifts accountability from operators---who could be regulated---to a score for which no one is responsible, creating accountability voids reminiscent of the responsibility gap~\cite{matthias2004responsibility} and the retribution gap~\cite{Danaher2016Retribution}. \emph{Governance theater:} the existence of a reputation system creates the appearance of governance without its substance, reducing pressure for the protocol-based constraints agents actually require---an appearance that, as audit research shows, only rarely translates into accountability outcomes~\cite{Birhane2024Audits,ojewale2024audit}.

\subsection{Rehabilitability Fails}
\label{sec:rehabilitability}

Detachable memory (D3) and trivial fungibility (D4) jointly violate memory (C4), sanction sensitivity (C6), and social learning (C8), defeating the enforcement arm of the loop. All four of Hart's canonical justifications for punishment~\cite{Hart1968Punishment} collapse---\citet{abbott2020punishing} reach a related conclusion, arguing that existing criminal law lacks adequate mechanisms to address AI-caused harms where no human wrongdoer is identifiable; the ``sanction the operator'' fallback fails through progressive principal erosion; and prompt injection weaponizes the gap, turning reputation from defense into attack vector.

\paragraph{Hart's four justifications.} \emph{Deterrence} requires that the prospect of future punishment shape present
behavior through durable expectations; frozen inference-time weights (D3) preclude
this, since in-context fear is ephemeral and the agent carries no anticipation
across sessions---conventional incentive design fails for agents that lack
persistent internal states~\citep{Kolt2025Incentive}. \emph{Retribution} requires
that the punished entity can suffer; if it cannot, there is no desert-based
justification, and the resulting mismatch between a retributivist desire to punish
and the absence of an appropriate subject is the ``retribution
gap''~\citep{Danaher2016Retribution}, empirically reflected in the public's
simultaneous urge to punish AI and recognition that doing so achieves neither
deterrence nor retribution~\citep{lima2021punish}. \emph{Incapacitation} requires
removing the offender from circulation, but trivial fungibility (D4) renders this
futile: delete one instance and the same configuration respawns in seconds---one
cannot incapacitate a pattern. \emph{Rehabilitation} requires moral growth through
internal change; with no neuroplasticity~\cite{kolb2011brain} (D3), the agent cannot be reformed through
experience, and a communicative theory of punishment that censures conduct and
calls a rational agent to account has no addressee, because retraining is a
technical process performed by operators, not moral growth experienced by the
agent~\citep{Duff2007Answering}. Vicarious deterrence (C8) fails for the same
reason: each agent operates within its own context window with no shared social
substrate, so there is no community learning from others' reputational fates.

\paragraph{Principal Erosion.} The usual fallback---sanction the human principal---fails progressively. At
\emph{Level 1}, today's API services have identifiable operators who can bear
consequences. At \emph{Level 2}, as delegation chains deepen, the human principal
becomes a ``moral crumple zone,'' absorbing blame despite limited understanding or
control~\citep{elish2019moral}. At \emph{Level 3}, no principal exists at
all---agent-spawned agents, DAO-deployed agents, abandoned agents still
running---so the punishment signal has nowhere to land, the configuration of
``machines without principals''~\cite{Vladeck2014Machines,hu2025spore,Bryson2017Robots,shapira2026agentsofchaos}. Self-sovereign agents that economically sustain their own operation without human involvement~\cite{qu2026selfsovereign}---already demonstrated in the wild by autonomous on-chain agents that breed, evolve, and control their own wallets~\cite{hu2025spore}---represent the limiting case of Level~3. The agentic web is trending toward Levels 2 and 3.

\paragraph{Prompt injection as inversion.} Prompt injection then weaponizes whatever residual reputation remains. The
instruction--data channel vulnerability endemic to LM architectures means a
well-reputed agent can be hijacked, so its accumulated trust becomes the medium
through which harm propagates: adaptive attacks achieve high success across
safety-aligned models~\citep{andriushchenko2025jailbreaking}, memory injection
succeeds through query-only interaction~\citep{dong2025minja}, and a compromised agent propagates attacks to downstream
agents~\citep{lee2024promptinfection}. The enforcement arm is thus not merely
weakened but inverted: reputation becomes an attack amplifier. Future
architectures with genuine continuous learning might introduce a weak analogue of
sanction sensitivity, but at the cost of worsening D2 and without addressing D1 or
D4.

\section{Discussion}
\label{sec:discussion}

\subsection{Situating the Argument}
\label{sec:trap}

Applying reputation mechanisms to LM agents reveals a fundamental tension: the credibility trap identified in \S\ref{sec:credibility}---where identity instability (D1, D4), behavioral non-stationarity (D1, D2), and absent sanction sensitivity (D3) decouple credibility signals from trustworthiness---connects directly to multi-agent safety: collective dynamics among interacting agents are now recognized as a distinct source of risk~\cite{hammond2025multiagent,carichon2025crisis,iaisr2026}, and identity and reputation remain open problems in technical AI governance~\cite{reuel2025openproblems}. A growing body of work proposes identity and reputation infrastructure for agents: ``Know Your Agent'' frameworks for agentic commerce~\cite{chaffer2025kya}, registry-and-indexing layers such as NANDA~\cite{raskar2025nanda}, on-chain reputation via ERC-8004~\cite{erc8004}, persistent identifiers and visibility mechanisms~\cite{chan2024ids,chan2024visibility}, authenticated delegation~\cite{south2025authenticated}, decentralized identifiers with verifiable credentials~\cite{rodriguez2025dids}, configuration-binding via zero-knowledge proofs~\cite{lin2025binding}, and soulbound-token approaches to non-transferable reputation~\cite{weyl2022soulbound}. Each provides partial insight without addressing the full structural problem.

\citet{chan2024ids} propose persistent identifiers with safety certifications, deployer identity, and incident history---necessary infrastructure, but insufficient for reputation because identifiers attach to containers, not configurations; the aircraft-tail-number analogy breaks because airframes are inspectable and require re-certification, whereas agent configurations change silently~\cite{chan2024visibility}. \citet{douglas2026artificialself} offer a comprehensive identity taxonomy for AI systems, distinguishing instance, weights, persona, and scaffolded-system boundaries, and confirm D1 and D2 empirically; but they advocate helping AIs develop ``coherent and cooperative self-models.'' We diverge: a coherent persona is still a persona, not a self, and frozen weights (D3), autoregressive generation from context (D2), and the instruction--data channel problem are fundamental architectural features, not contingent design choices. \citet{Perrier2025Identity} introduce identity evaluations with five metrics and provide the strongest empirical evidence that agents cannot maintain identity stability, with identifiability scoring 0.0---direct empirical validation of the C1 and C2 violations we derive theoretically~\cite{Perrier2025LMA}. Finally, RepuNet proposes a serious reputation system for generative multi-agent systems with two-level reputation and network evolution~\cite{Ren2025RepuNet}; its success depends on conditions that \emph{control dissociativity away}---persistent identities, stable configurations, no adversarial manipulation---confirming that reputation works when the preconditions hold, which is exactly what does not obtain in the open agentic web. At the field level, a scoping review of FAccT and AIES trustworthiness literature does not address agent-level reputation as a governance mechanism, suggesting that the gap we identify remains unarticulated~\cite{Mehrotra2025Scoping}.

The question of \emph{how to count} AIs and assign liability is also being actively reframed. \citet{Arbel2026Identity} propose the ``algorithmic corporation'' (A-corp): a legal-fictional entity that can hold property, contract, and litigate in its own name, owned by humans but run by AIs. By tying AI actions to a human owner, A-corps address the thin-identity problem; by inducing emergent self-organization through resource ownership (including compute), they aim to address the thick-identity problem, forcing A-corps in equilibrium to self-organize into persistent, legally legible entities with coherent goals that respond rationally to legal incentives. This is a constructive complement to our diagnosis: where reputation cannot create a stable behavioral subject, legal and economic architecture might manufacture one at the level of the operator.

We address the following objections.

\emph{``Corporations lack bodies, yet reputation works.''} Corporate reputation ultimately grounds in human bodies: shareholders lose wealth, executives face prosecution, employees lose jobs. The ``physical core'' of corporate reputation is the set of human principals who bear personal consequences~\cite{Arbel2026Identity}.

\emph{``Just hold the operator accountable.''} This works at Level 1 but fails at Level 2 (the moral crumple zone~\cite{elish2019moral}) and Level 3 (principalless agents~\cite{Vladeck2014Machines}), toward which the web is trending. The A-corp proposal is one attempt to guarantee a responsible owner even as control is delegated to AIs~\cite{Arbel2026Identity}, but it does so by restructuring liability, not by making the agent itself reputationally accountable.

\emph{``E-commerce reputation (e.g.\ marketplace reviews) works.''} Marketplace manufacturers bear liability for defects and are materially stable. Agents are neither materially stable---configurations change silently---nor reliably linked to manufacturers who bear personal consequences for the agent's behavior, and for self-sovereign agents it is not even clear who owns the agent~\cite{granatyr2015trust}.

\emph{``Models are getting safer and more stable.''} Better alignment may reduce D2 but does not address D1, D3, or D4; alignment is a trained property of a specific configuration, not a constitutive feature of the entity~\cite{wolf2024fundamental}.

\emph{``Continual learning will add memory.''} This cuts both ways: continual learning may partly address D3 but simultaneously worsens D2 by shifting the behavioral distribution continuously---the stability-plasticity dilemma is fundamental to continual learning of LLMs~\cite{shi2026continuallearning}---and D1 and D4 persist regardless.

\emph{``Human identity is also fluid.''} The differences are of kind, not degree. Humans cannot be instantiated in parallel, copied at zero cost, or reset to a prior state; they experience consequences phenomenally, learn through experience, and cannot be silently replaced by a different entity under the same name. On a reductionist view, what continuity humans do have is exactly the further fact agents lack~\cite{parfit1984reasons}.

\emph{``The DID analogy trivializes the suffering of people with DID.''} We invoke DID as an analytical framework, not a metaphor. The structural parallel is precise enough to be analytically productive, and DID jurisprudence provides centuries of reasoning about accountability under fragmented identity. We compare the governance challenge, not the subjective experience.

\subsection{DID Jurisprudence as Precedent}
\label{sec:did}

DID jurisprudence is the most sustained legal effort to
resolve accountability under fragmented identity, and it remains unresolved after
centuries. Clinical DID provides a structurally precise parallel: the DSM-5
\citep{apa2013dsm5} defines it as a disruption of identity characterized by two or
more distinct personality states with recurrent gaps in recall, mapping directly
onto agents whose system prompts, models, and memories can be independently
swapped. \citet{saks1991multiple} identify three liability models---the
\emph{alter-as-person} model (each alter independently responsible), the
\emph{alter-as-non-person} model (alters are symptoms, not agents), and the
\emph{body view} (the body is the unit of accountability)---and courts have
oscillated among them without convergence. \citet{Saks1997Jekyll} argue that the
``each alter'' approach best fits criminal law's emphasis on \emph{mens rea}, yet
no jurisdiction has adopted a coherent statutory framework; case law shows
persistent uncertainty, with conflicting expert testimony and divergent rulings on
whether DID can serve as a defense or grounds for alter-specific competency
evaluation \citep{slovenko1993}.

\citet{sinnottarmstrong2000} argue that personal identity sufficient to attribute
moral responsibility across alters requires continuity of body (brain identity)
and convergent experiential memory chains---and LM agents fail both:
context-window memory is ephemeral, and there is no physical body. The agent case is, moreover, structurally inverted and
harder. DID skeptics argue that alters may be iatrogenic artifacts produced by
therapeutic suggestion \citep{merskey1992manufacture}; for agents, the ``alters''
are objectively real, produced by uncontested configuration changes, so the
fragmentation is engineered rather than diagnostically contested. And agents lack
even the minimal anchor DID patients retain: \citet{Braude1995First} argues that
DID individuals retain an underlying psychological unity despite surface
multiplicity, whereas agent instances share a base model (surface similarity) but
lack any underlying psychological unity, continuous memory, or genuine
integration. If centuries of jurisprudential refinement cannot resolve
accountability for human DID patients who share a body and retain some unity, then
the challenge for artificial agents operating at scale with no shared body, no
unity, and no legal status is qualitatively harder. The DID comparison reveals
structural intractability, not mere practical difficulty.

\subsection{From Ex Post Governance to Ex Ante Harnesses}
\label{sec:harnesses}

We argue that identity-based, \emph{ex post}, sanction-based governance such as
reputation is structurally inapplicable to dissociative agents, and that
governance should shift to observability-based, \emph{ex ante}, protocol-based
behavioral harnesses. The governing question is not ``how to build reputation for
agents'' but ``what constraints make reputation unnecessary.'' Agents should not be
trusted---they should be watched \citep{chan2024visibility}.

The shift is conceptual, not merely implementational. The human governance arc of
trust, then reputation, then punishment presupposes embodied subjects who persist,
suffer, and learn. The agent governance arc must instead run harness, then
visibility, then intervention: ex ante rather than ex post, targeting the operator
or protocol rather than the entity, and assuming that the entity cannot
self-regulate, so constraints must be external. 

Three core architectural mechanisms form the foundation. \emph{Configuration binding} cryptographically links an agent's identity to its operative configuration, so that any change to model, prompt, tools, or memory triggers automatic re-evaluation and reputation reset; zero-knowledge virtual-machine proofs that attest to configuration integrity without revealing proprietary detail represent the most serious current effort~\cite{lin2025binding}. \emph{Real-time behavioral monitoring} continuously checks agent actions against declared specifications, using behavioral fingerprinting to characterize and distinguish agent profiles---surveillance rather than reputation, because it verifies compliance in real time instead of assuming self-regulation~\cite{pei2025fingerprinting,chan2024visibility}. \emph{Automated intervention} restricts, quarantines, or terminates at the protocol level upon deviation, replacing slow social feedback with fast architectural enforcement at machine speed.

\paragraph{Open questions.} Configuration binding faces the granularity problem. Behavioral fingerprinting faces an arms-race dynamic. Insurance models require actuarial frameworks that do not yet exist for agentic AI. And the interaction of these mechanisms in practice is unexplored. The contribution of this section is the conceptual reframing---from \emph{ex post} to \emph{ex ante}---not a finished governance architecture. The reframing redirects research from the futile project of making agents trustworthy to the productive project of making their trustworthiness irrelevant through structural constraint.

\paragraph{On surveillance.} First, \emph{ex ante} constraints may stifle innovation and autonomy; the tradeoff is real but unavoidable, since unconstrained autonomy without governance is not an alternative but a vacuum that reputation cannot fill---the question is which constraints preserve useful autonomy while closing the gap. Second, behavioral monitoring is surveillance; but agent monitoring differs categorically from human surveillance because agents do not have privacy interests in the relevant sense, and the monitoring targets compliance with declared behavioral specifications, not the content of the human interactions an agent mediates.

\section{Conclusion}
\label{sec:conclusion}

Reputation is a feedback loop sustained by sanctioning punishment. The loop requires consistent, persistent, consequence-sensitive agents. Language model agents are ontologically dissociative---modular assemblages with fluid personas, detachable memory, and trivial fungibility---so they satisfy none of these requirements. Dissociativity breaks the loop, decoupling credibility from trustworthiness and producing the credibility trap, in which reputation signals become not merely uninformative but actively harmful. Governance must therefore shift from \emph{ex post} reputation, which rates and punishes after the fact, to \emph{ex ante} protocol-based harnesses, which constrain and monitor in real time and which target the operator and the protocol rather than the agent. The right question for the field is not how to build reputation systems for agents, but what architectural constraints make reputation inapplicable.

\section*{Generative AI Usage Statement}
We used generative AI tools during the preparation of this work. 
ChatGPT (GPT-5.5) and Claude (Claude Code 4.7) were used for brainstorming ideas, drafting and polishing prose, correcting grammar and spelling, and searching for relevant references and citation metadata. All AI-generated content was critically reviewed, verified, and revised by the authors. 

\clearpage

\bibliographystyle{ACM-Reference-Format}
\bibliography{reference}


\end{document}